# Closed spherically symmetric massless scalar field spacetimes have finite lifetimes

Gregory A. Burnett

*Department of Mathematics, North Carolina State University, Raleigh, NC 27695-8205*

(April 26, 1994)

The closed-universe recollapse conjecture is studied for a class of closed spherically symmetric spacetimes which includes those having as a matter source: (1) a massless scalar field; (2) a perfect fluid obeying the equation of state $\rho = P$; and (3) null dust. It is proven that all timelike curves in any such spacetime must have length less than $6 \max_\Sigma (2m)$, where $m$ is the mass associated with the spheres of symmetry and $\Sigma$ is any Cauchy surface for the spacetime. The simplicity of this result leads us to conjecture that a similar bound can be established for the more general spherically symmetric spacetimes.

04.20.Dw, 04.20.Cv, 98.80.Hw

## I. INTRODUCTION

It has been conjectured [1–4] that if our Universe has the spatial topology of a three-sphere ($S^3$) or a three-handle ($S^1 \times S^2$) and the dominant matter content is "ordinary" (e.g., all principal pressures are non-negative and no greater than the energy density) then, eventually, its expansion will cease (at which time the Universe has reached its maximal size) and thereafter it will recollapse back on itself. It is well known that such behavior occurs in the spatially homogeneous and isotropic models with $S^3$ spatial topology (i.e., the $k = +1$ Robertson-Walker spacetimes) with dust or radiation as a matter source. The *closed-universe recollapse conjecture* asserts that the behavior seen in these simple models is a generic feature of general relativity.

An alternate (weaker) version of the closed-universe recollapse conjecture asserts that if our Universe meets the conditions above, it will have a finite lifetime in the sense that there will be a finite upper bound to the lifetimes of all observers therein [4–6]. There is no claim that the Universe will actually reach a maximal size and then recollapse. One precise version of this conjecture is the following [5,6].

**Conjecture:** There exists an upper bound to the lengths of timelike curves in any spacetime that possesses $S^3$ or $S^1 \times S^2$ Cauchy surfaces and that satisfies the dominant-energy and non-negative-pressures conditions.

Herein, the dominant-energy condition is the demand on the Einstein tensor $G_{ab}$ that $G_{ab} t^a u^b \geq 0$ for all future-directed timelike $t^a$ and $u^b$, while the non-negative-pressures condition is the demand that $G_{ab} x^a x^b \geq 0$ for all spacelike $x^a$. For example, these conditions are satisfied for a perfect fluid spacetime with energy density $\rho$ and pressure $P$ if and only if $0 \leq P \leq \rho$.

Why should one believe that such a conjecture should be true? In addition to the fact that there is no known counterexample, there are results in its favor. It has been proven that the conjecture holds for the spatially homogeneous spacetimes: those with $S^3$ Cauchy surfaces being the Bianchi type IX spacetimes [7]; and those with $S^1 \times S^2$ Cauchy surfaces being the Kantowski-Sachs spacetimes [5]. It has also been proven that the conjecture holds for the spherically symmetric spacetimes with $S^1 \times S^2$ Cauchy surfaces. While its status for those with $S^3$ Cauchy surfaces is presently unknown, recently it has been proven [6] that those spherically symmetric spacetimes having dust as a matter source (the Tolman spacetimes) do satisfy the above conjecture.

Here, with an eye towards resolving the above conjecture for the spherically symmetric spacetimes with $S^3$ Cauchy surfaces, we investigate its validity for those spherically symmetric spacetimes satisfying the additional condition that the trace of that part of the Einstein tensor perpendicular to the spheres of symmetry be zero. Denoting that part of the Einstein and metric tensors perpendicular to the spheres of symmetry by $\tau^{ab}$ and $h^{ab}$ respectively, this condition can be written as $G_{ab} h^{ab} = \tau_{ab} h^{ab} = \tau^a{}_a = 0$. Our restriction to these spacetimes arises from two requirements. First, the dominant-energy condition requires that $\tau^a{}_a$ be non-positive. [Proof: Using the fact that $h^{ab} = -2k^{(a} l^{b)}$ where $k^a$ and $l^a$ are any two null vectors perpendicular to the spheres of symmetry such that $k_a l^a = -1$, we have $\tau_{ab} h^{ab} = -2\tau_{ab} k^a l^b \leq 0$, where the inequality follows by the dominant-energy condition.] Second, when $\tau^a{}_a$ is non-negative, then $D^a D_a r$ is non-negative. (See Eq. (2.6).) This is a technical requirement needed for the method of proof used herein. Clearly, these requirements are both satisfied if and only if $\tau^a{}_a = 0$. While this condition severely limits the class of spacetimes being investigated, it does include three interesting cases:

*(1) Massless scalar field spacetimes.* Here the matter field is a (spherically symmetric) scalar field $\phi$. In this case,

$$G_{ab} = 8\pi \left[ \nabla_a \phi \nabla_b \phi - \frac{1}{2} g_{ab} (\nabla_m \phi \nabla^m \phi) \right]. \quad (1.1)$$

Clearly, $\tau^a{}_a = G_{ab} h^{ab} = 0$. The dominant-energy condition is always satisfied.

*(2) Perfect fluid spacetimes with $\rho = P$.* A perfect fluid is described by a unit-timelike vector $u^a$, and two scalar fields $\rho$ (the energy density) and $P$ (the pressure). In this case,

$$G_{ab} = 8\pi \left[ (\rho + P) u_a u_b + P g_{ab} \right]. \quad (1.2)$$

Again, we find that $\tau^a{}_a = G_{ab} h^{ab} = 8\pi(P - \rho) = 0$. The dominant-energy condition is satisfied iff both $\rho$ and $P$ (being equal) are non-negative.

*(3) Null dust spacetimes.* Here the matter field consists of two scalar fields $\rho_1$ and $\rho_2$ and their associated radial (non-colinear) null vector fields $k_1^a$ and $k_2^a$ (which can be chosen to be geodetic). In this case,

$$G^{ab} = 8\pi \left[ \rho_1 k_1^a k_1^b + \rho_2 k_2^a k_2^b \right], \quad (1.3)$$

which satisfies the condition that $\tau^a{}_a = 0$ as $k_1^a$ and $k_2^a$ are both null. The dominant-energy condition is satisfied iff $\rho_1, \rho_2 \geq 0$.

Our result concerning these spacetimes is summarized by the following theorem.

**Theorem 1.** The length of any timelike curve in a spherically symmetric spacetime possessing a compact Cauchy surface $\Sigma$, satisfying the dominant-energy condition, and satisfying the condition that $G_{ab} h^{ab} = 0$ must be less than $6 \max_\Sigma (2m)$, where $m$ is the mass associated with the spheres of symmetry.

This result and the methods used in its proof are promising in many respects. First, the bound given in theorem 1 is both beautifully simple (compare the appearance of this expression to Eq. (2.11) of Ref. [5] and Eq. (3.17) of Ref. [6]) and usually much smaller than those found before. (It can be larger, though never more than by a factor of $6/\pi$.) Second, the method used to establish theorem 1 takes very little advantage of the "specialness" of the spacetimes being considered (i.e., the condition that $\tau^a{}_a = 0$). This is to be compared to the analogous result for the Tolman spacetimes [6] where a number of the properties of the Tolman spacetimes were used. This suggests that, with some further insight, the methods used here may be modified to prove the more general case. Third, although the method of proof here is apparently dissimilar to the proof of the Tolman result, they both use a similar quantity—the derivative of $r$ along certain null directions. This similarity may be a coincidence, but for now it appears promising. Fourth, the case considered here and the Tolman case are, in a sense, the two extreme cases for the spherically symmetric spacetimes. If we restrict ourselves to the perfect-fluid spherically symmetric spacetimes, the dominant-energy and non-negative-pressures conditions require that $0 \leq P \leq \rho$. In the Tolman case, the pressure is minimal: $P = 0$. In the case considered herein, the pressure is maximal: $P = \rho$. Having proofs for the two extreme cases suggest that a proof for the intermediate cases can be found. Fifth, and last, the proof used herein works equally well in the $S^3$ and $S^1 \times S^2$ cases. This suggest that the methods used are probably close to "the way things should be done".

One might wonder why a pressure condition was not imposed in theorem 1 as was done in prior works on the closed-universe recollapse conjecture. In the case here, as a consequence of the dominant-energy condition and the condition that $\tau_{ab} h^{ab} = 0$, the *radial-non-negative-pressures condition* is automatically satisfied. This is the condition that $G_{ab} x^a x^b \geq 0$ for all spacelike vectors $x^a$ perpendicular to the spheres of symmetry. [Proof: At each point choose two radial non-colinear null vectors $k^a$ and $l^a$. Then for some scalars $\alpha$, $\beta$, and $\gamma$ we can write $\tau_{ab} = \alpha k_a k_b + \beta l_a l_b - \gamma h_{ab}$. The dominant-energy condition demands that $\alpha, \beta, \gamma \geq 0$ while the $\tau_{ab} h^{ab} = 0$ condition requires that $\gamma = 0$. Therefore, $\tau_{ab}$ is positive-indefinite: $\tau_{ab} \xi^a \xi^b \geq 0$ for all $\xi^a$, in particular radial-spacelike $\xi^a$.] This pressure condition is sufficient for the work herein, and it is interesting to note that this condition would have been sufficient for the results in Refs. [5,6]. (The radial-non-negative-pressures condition is somewhat unsatisfactory in the sense that its natural generalization to arbitrary spacetimes is the non-negative-pressures condition. However, imposing that condition here would exclude the massless scalar field spacetimes.)

The proof of theorem 1 can be summarized as follows. As is argued in Sec. III, it is sufficient to establish theorem 1 for the two timelike geodesics $\gamma_n$ and $\gamma_s$ along which $r = 0$ (i.e., the world lines of the two observers for whom the universe actually appears spherically symmetric—existing only in the $S^3$ case), and radial timelike geodesics which don't intersect $\gamma_n$ or $\gamma_s$ (excepting possibly at its endpoints). To this end, it was noted that for the spacetimes being considered, $D^a D_a r \geq 0$. Integrating this inequality over the "triangle" formed by a segment of $\gamma_n$ or $\gamma_s$ and two radial null curves and then using the fact that $r$ is bounded above immediately gives a simple upper bound for the length of the segment and hence $\gamma_n$ and $\gamma_s$. That we can show that the curves $\gamma_n$ and $\gamma_s$ have finite length so simply is remarkable. Arguments in prior works that worked so well in the $S^1 \times S^2$ case failed in the $S^3$ case because of the existence of the curves $\gamma_n$ and $\gamma_s$.

Although establishing a bound on the lengths of $\gamma_n$ and $\gamma_s$ is remarkably easy, bounding the lengths of the radial timelike geodesics is, by comparison, awkward. That an argument could be found for these curves was motivated by the following observations. For curves "near" $\gamma_n$ or $\gamma_s$ one can repeat the argument used above to establish an upper bound on their lengths. For radial timelike geodesics "far" from $\gamma_n$ and $\gamma_s$ in the sense that $2m/r \geq 2\epsilon > 0$ all along the curve, then one has $\ddot{r} \leq -\epsilon/r$. (See Eq. (2.5) below.) From this and the fact that $r$ is bounded above (theorem 2 below) it follows that such a curve is bounded in length. Therefore, since one can bound the lengths of curves in the two extreme cases, it would seem that by suitably combining these arguments a bound on the lengths of all radial timelike curves could be achieved. This is done in Sec. III C.

In Sec. II, the basics of the spherically symmetric spacetimes are briefly reviewed. In Sec. III, the full details of the proof of theorem 1 are given. Lastly, in Sec. IV, a few final remarks are made regarding possible extensions of this work.

The conventions used herein are those of Ref. [8]. In particular, our metrics are such that timelike vectors have negative norm and the Riemann and Ricci tensors are defined by $2\nabla_{[a} \nabla_{b]} \omega_c = R_{abc}{}^d \omega_d$ and $R_{ab} = R_{amb}{}^m$ respectively. All metrics are taken to be $C^2$. Our units are such that $G = c = 1$.

## II. REVIEW

In this section, the basic features of the spherically symmetric spacetimes needed here are reviewed. For a more complete presentation, see Refs. [5] and [6].

Recall that a spacetime $(M, g_{ab})$ is said to be *spherically symmetric* if it admits a group $G \approx SO(3)$ of isometries, acting effectively on $M$, each of whose orbits is either a two-sphere or a point [9]. Denote the orbit of a point $p$ by $\mathcal{S}_p$. The value of the non-negative scalar field $r$ at each $p \in M$ is defined so that $4\pi r^2$ is the area of $\mathcal{S}_p$. So, in particular, $r(p) = 0$ if $\mathcal{S}_p = p$, while $r(p) > 0$ if $\mathcal{S}_p$ is a two-sphere. In the case where $(M, g_{ab})$ has $S^3$ Cauchy surfaces, the set of points where $r = 0$ consists of two disconnected geodesics which we label as $\gamma_n$ and $\gamma_s$.

Where $r > 0$, we decompose the metric $g_{ab}$ into the sum $g_{ab} = h_{ab} + q_{ab}$, where $q^a{}_b$ is the projection operator onto the tangent space of each sphere of symmetry and $h^a{}_b$ is the projection operator onto the tangent space of each two-surface perpendicular to the spheres of symmetry. Using the fact that there exists a preferred "unit-metric" $\Omega^{ab}$ on each sphere of symmetry, we have $q_{ab} = r^2 \Omega_{ab}$ (where $\Omega^{am}\Omega_{mb} = q^a{}_b$ and $\Omega_{ab} = q^m{}_a q^n{}_b \Omega_{mn}$). This gives us the final decomposition of $g_{ab}$ as

$$g_{ab} = h_{ab} + r^2 \Omega_{ab}. \tag{2.1}$$

In addition to the derivative operator $\nabla_a$ associated with metric $g_{ab}$, we have the derivative operator $D_a$ associated with the (unphysical) metric $h_{ab} + \Omega_{ab}$. As argued in Ref. [6], $D_a$ can be thought of as the derivative operator associated with $h_{ab}$ on the surfaces perpendicular to the spheres of symmetry ($D_a h_{bc} = 0$) and as the derivative operator associated with $\Omega_{ab}$ on the spheres of symmetry ($D_a \Omega_{bc} = 0$).

For the spherically symmetric spacetimes, the mass $m$ associated with the spheres of symmetry is defined by

$$2m = r(1 - D_m r D^m r). \tag{2.2}$$

The two most important properties of the scalar fields $r$ and $m$ are summarized by the following theorem. (For a proof, see Refs. [5,6]. Note that although the theorems as stated in these references demand that the non-negative-pressures condition hold, they need only demand the radial-non-negative-pressure condition.)

**Theorem 2.** For any spherically symmetric spacetime that possesses a compact Cauchy surface $\Sigma$ and that satisfies the dominant-energy and radial-non-negative-pressure conditions, $m \geq 0$ and $r \leq \max_\Sigma (2m)$.

Next, defining $\epsilon^{ab}$ to be either of the two antisymmetric tensor fields such that $\epsilon^{ab}\epsilon^{cd} = -2h^{a[c}h^{d]b}$, and denoting the "radial part" of the Einstein tensor $G^{ab}$ by $\tau^{ab}$ (i.e., $\tau^{ab} = h^a{}_m h^b{}_n G^{mn}$) we have

$$D_a D_b r = \frac{m}{r^2} h_{ab} - \frac{r}{2}\tau^{mn}\epsilon_{ma}\epsilon_{nb}. \tag{2.3}$$

For the spacetimes being considered, $\tau^{mn}\epsilon_{ma}\epsilon_{nb} = \tau_{ab} - \tau^m{}_m h_{ab} = \tau_{ab}$. So Eq. (2.3) becomes

$$D_a D_b r = \frac{m}{r^2} h_{ab} - \frac{r}{2}\tau_{ab}. \tag{2.4}$$

From Eq. (2.3) we shall need only two facts. First, along any radial timelike geodesic with unit-tangent vector $t^a$ and parameter $t$

$$\frac{d^2 r}{dt^2} \leq -\frac{m}{r^2}, \tag{2.5}$$

which follows by contracting Eq. (2.3) with $t^a t^b$ and the radial-non-negative-pressure condition. Second, contracting Eq. (2.3) with $h^{ab}$ we find that

$$D^a D_a r = \frac{2m}{r^2} + \frac{r}{2}\tau^a{}_a \geq 0, \tag{2.6}$$

where the inequality follows from the non-negativity of $m$ and the fact that $\tau^a{}_a = 0$ for the spacetimes being considered.

## III. PROOF OF THEOREM 1

Recall that the distance function $d(p,q)$ is defined for $q \in I^+(p)$ to be the least upper bound to the lengths of timelike curves from $p$ to $q$ (and is defined to be zero otherwise) [9]. Thus, a finite upper bound to the lengths of timelike curves in a spacetime will exist if and only if $d(p,q)$ is bounded above by some constant independent of $p$ and $q$.

Fix $p, q \in M$ with $q \in I^+(p)$ and any timelike curve $\mu$ connecting $\mathcal{S}_p$ to $\mathcal{S}_q$ having length $d(\mathcal{S}_p, \mathcal{S}_q)$. Such a curve $\mu$ is necessarily geodetic, maximal, radial (perpendicular to spheres of symmetry), and has a length no less than $d(p,q)$. From the maximality of $\mu$ it follows (see Appendix A) that either $\mu$ is a segment of $\gamma_n$ or $\gamma_s$ (i.e., $r = 0$ on $\mu$) or only the endpoints of $\mu$ can possibly intersect $\gamma_n$ or $\gamma_s$ (i.e., $r > 0$ between the endpoints of $\mu$). Therefore, to prove theorem 1, we need only bound $d(\mathcal{S}_p, \mathcal{S}_q)$ and we do this by bounding the lengths of $\gamma_n$, $\gamma_s$, and radial timelike geodesics on which $r > 0$ (excepting possibly its endpoints).

To bound the lengths of such curves, we first establish three key inequalities in Sec. III A. We then apply them in Sec. III B to bound the lengths of $\gamma_n$ and $\gamma_s$ and then again in Sec. III C to bound the lengths of radial timelike geodesics on which $r > 0$ (excepting possibly endpoints.)

### A. Three key inequalities

For each one-form $\omega_a$, we define the one-form $^*\omega_a$ by setting

$$^*\omega_a = \omega_b \epsilon^b{}_a. \tag{3.1}$$

While this is not quite the Hodge-dual [10] of a one-form on $M$ (as this is a 3-form), it can be thought of as the Hodge-dual of one-forms on the surfaces perpendicular to the spheres of symmetry.

Along any timelike curve $\gamma$, parameterized by $t$ so that its tangent vector $t^a$ is unit-timelike ($t^a t_a = -1$), define the quantity $Q$ by

$$Q = -t^a (^*dr)_a. \tag{3.2}$$

Although $(dr)_a$ and $\epsilon^{ab}$ are discontinuous on the curves $\gamma_n$ and $\gamma_s$, the combination $(^*dr)^a$ is well-behaved there in the sense that it admits a continuous extension to these curves (being a unit tangent vector to these curves under such an extension). Note that $Q$ depends on the choice of $\epsilon^{ab}$. However, in the case where $\gamma$ is a segment of $\gamma_n$ or $\gamma_s$, we shall take $\epsilon^{ab}$ so that $(^*dr)^a$ is parallel to $t^a$. With this choice, $Q = +1$ on such curves.

On a radial timelike curve $\gamma$ that does not intersect $\gamma_n$ or $\gamma_s$ (except possibly at its endpoints), define

$$x^a = {}^*t^a. \tag{3.3}$$

Note that $x^a$ is radial, unit-spacelike ($x^a x_a = +1$), and orthogonal to $t^a$ ($t^a x_a = 0$). Furthermore, $Q = x^a(dr)_a$, i.e., $Q$ is the "spatial derivative" of $r$ as measured by an observer whose world line is $\gamma$. Further define $\pm k^a = \pm t^a + x^a$, i.e., $\pm k^a$ are the two radial null vectors on the "+$x$-side" of $\gamma$ having unit-spacelike part ($\pm k^a x_a = +1$). On $\gamma$, we define $Q^+$ and $Q^-$ to be the derivatives of $r$ along these null directions:

$$Q^\pm = \pm k^a (dr)_a, \tag{3.4a}$$
$$= \pm \dot{r} + Q, \tag{3.4b}$$

where $\dot{r} = t^a (dr)_a$. (Although we shall not need $Q^\pm$ to be defined on $\gamma$ in the case where it is a segment of $\gamma_n$ or $\gamma_s$, in view of Eq. (3.4b), any reasonable definition should give $Q^\pm = 1$.)

Defining

$$r_M = \max_\Sigma (2m), \tag{3.5}$$

where $\Sigma$ is any Cauchy surface for the spacetime, we have the following lemma.

**Lemma 1.** Fix any timelike curve $\gamma$ from $p_1 = \gamma(t_1)$ to $p_2 = \gamma(t_2)$ that is either a segment of $\gamma_n$ or $\gamma_s$ or is radial and does not intersect $\gamma_n$ or $\gamma_s$ (except possibly at $p_1$ or $p_2$). Then,

$$\int_{t_1}^{t_2} Q \, dt \leq 2r_M - r(p_1) - r(p_2), \tag{3.6a}$$

$$\int_{t_1}^{t_2} Q^+ \, dt \leq 2(r_M - r(p_1)), \tag{3.6b}$$

$$\int_{t_1}^{t_2} Q^- \, dt \leq 2(r_M - r(p_2)). \tag{3.6c}$$

*Proof.* Using the relation between $Q$ and $Q^\pm$ given by Eq. (3.4b), it is straightforward to show that the above three inequalities are all equivalent, so we establish all three by proving Eq. (3.6a). To this end, consider the compact set

$$K = J^+(\mathcal{S}_{p_1}) \cap J^-(\mathcal{S}_{p_2}). \tag{3.7}$$

In the case where $\gamma$ is a segment of $\gamma_n$ or $\gamma_s$, define $C$ to be the two-dimensional (compact) region that is the intersection of $K$ and any of the two-surfaces perpendicular to the spheres of symmetry. (For the purposes of this construction, such two-surfaces are to include $\gamma_n$ and $\gamma_s$.)

In the case where $\gamma$ is a radial timelike curve that does not intersect $\gamma_n$ or $\gamma_s$ (excepting $p_1$ and $p_2$), the orbit $\mathcal{T}$ of $\gamma$ (a timelike three-surface) divides $K$ into two disconnected components. (Although this need not be true when $(M, g_{ab})$ has $S^1 \times S^2$ Cauchy surfaces, it is true for its universal covering spacetime which can be used in its place in this construction.) Define $C$ to be the two-dimensional (compact) region that is the intersection of that half of $K$ into which $x^a$ is inward pointing and that two-surface perpendicular to the spheres of symmetry which contains $\gamma$.

In each case, the boundary of $C$ will consist of either: (1) the timelike curve $\gamma$ from $p_1$ to $p_2$, a null geodesic $\nu_2$ from $p_2$ to a point $q$, and a null geodesic $\nu_1$ from $q$ to $p_1$; or (2) the timelike curve $\gamma$ from $p_1$ to $p_2$, a null geodesic $\nu_2$ from $p_2$ to a point $q_2$ (on either $\gamma_n$ or $\gamma_s$), a segment $\sigma$ of $\gamma_n$ or $\gamma_s$ from $q_2$ to $q_1$, and a null geodesic $\nu_1$ from $q_1$ to $p_1$. (See Fig. 1.) We deal with each case separately.

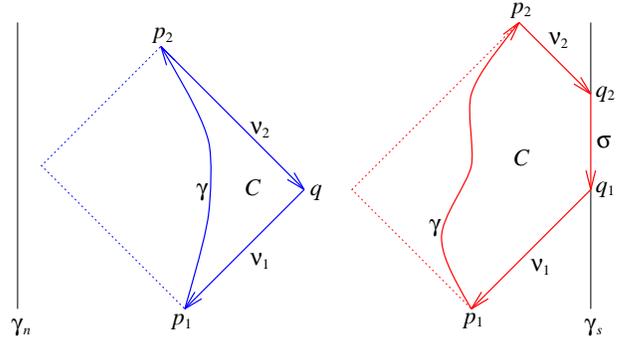

FIG. 1. A planar slice of $M$ (perpendicular to the spheres of symmetry) showing a radial timelike curve $\gamma$ connecting $p_1$ to $p_2$ and the compact region $C$ constructed therefrom in two different cases. The left curve corresponds to Case 1 while the right corresponds to Case 2. (Note: With the future being "upwards", the choice of $\epsilon^{ab}$ depicted here is such that $x^a$ is a "right-directed" vector along any future-directed timelike curve in this diagram.)

*Case 1.* Choosing the orientation of $C$ corresponding to $\epsilon_{ab}$ and the induced orientation of the boundary of $C$ to be that shown in Fig. 1 (i.e., the orientation corresponding to going from $p_1$ to $p_2$ to $q$ and back to $p_1$) then, by Stokes's theorem, we have

$$-\int_\gamma {}^*dr = -\int_C d({}^*dr) + \int_{\nu_2} {}^*dr + \int_{\nu_1} {}^*dr. \tag{3.8}$$

We now evaluate each term.

$$-\int_\gamma {}^*dr = \int_{t_1}^{t_2} [-t^a({}^*dr)_a] dt = \int_{t_1}^{t_2} Q \, dt, \tag{3.9a}$$

$$\int_C d({}^*dr) = \int_C (D^m D_m r)\epsilon \geq 0, \tag{3.9b}$$

$$\int_{\nu_2} {}^*dr = \int_{\nu_2} dr = r(q) - r(p_2), \tag{3.9c}$$

$$\int_{\nu_1} {}^*dr = -\int_{\nu_1} dr = r(q) - r(p_1). \tag{3.9d}$$

In Eq. (3.9a), the first equality follows from the definition of the line integral. Eq. (3.9b) follows from the fact that $[d({}^*dr)]_{ab} = (D^m D_m r)\epsilon_{ab}$ and Eq. (2.6). Eqs. (3.9c) and (3.9d) follow from the facts that for any vector $k_2^a$ tangent to $\nu_2$ and any vector $k_1^a$ tangent to $\nu_1$, ${}^*k_2^a = -k_2^a$ and ${}^*k_1^a = k_1^a$ so that $k_2^a({}^*dr)_a = -({}^*k_2)^a(dr)_a = k_2^a(dr)_a$, and similarly $k_1^a(^*dr)_a = -k_1^a(dr)_a$. Combining these results, it follows from Eq. (3.8) that

$$\int_{t_1}^{t_2} Q \, dt \leq 2r(q) - r(p_1) - r(p_2). \tag{3.10}$$

Eq. (3.6a) now follows from Eq. (3.10) and the fact that $r(q) \leq r_M$.

*Case 2.* Choosing the orientations as in Case 1, by Stokes's theorem, we have

$$-\int_\gamma {}^*dr = -\int_C d(^*dr) + \int_{\nu_2} {}^*dr + \int_{\nu_1} {}^*dr + \int_\sigma {}^*dr. \tag{3.11}$$

Repeating the arguments as before, we now find that

$$\int_{\nu_2} {}^*dr = -r(p_2) \leq 0, \tag{3.12a}$$

$$\int_{\nu_1} {}^*dr = -r(p_1) \leq 0, \tag{3.12b}$$

$$\int_\sigma {}^*dr = -(\text{length of } \sigma) \leq 0. \tag{3.12c}$$

In Eq. (3.12c), the first equality follows from the fact that $(^*dr)^a$ is a unit past-directed timelike vector on $\sigma$.

Combining these results, it follows from Eq. (3.11) that

$$\int_{t_1}^{t_2} Q \, dt \leq 0. \tag{3.13}$$

That Eq. (3.6a) holds in this case now follows from the fact that $r \leq r_M$. □

### B. Bound on lengths of $\gamma_n$ and $\gamma_s$

With lemma 1, a bound on the lengths of $\gamma_n$ and $\gamma_s$ is easily established.

**Theorem 3.**

$$(\text{length of } \gamma_n) \leq 2r_M, \tag{3.14a}$$

$$(\text{length of } \gamma_s) \leq 2r_M. \tag{3.14b}$$

*Proof.* Take $\gamma$ to be any connected segment of $\gamma_n$ or $\gamma_s$. Using the fact that $Q = +1$ on $\gamma$ and $r(p_1) = r(p_2) = 0$, Eq. (3.6a) immediately gives $2r_M$ as an upper bound for the length of such a segment. Since the choice of segment is arbitrary, Eq. (3.14) now follows. □

Note that this bound on the lengths of $\gamma_n$ and $\gamma_s$ is smaller (by a factor of 3) than that given by theorem 1.

### C. Bound on lengths of radial timelike geodesics

Having taken care of $\gamma_n$ and $\gamma_s$, we now complete the proof of theorem 1 by bounding the lengths of radial timelike geodesics upon which $r > 0$ (excepting possibly its endpoints). From Eq. (2.5) and the non-negativity of $m$, it follows that such a curve $\mu$ can be broken into the connected curves: $\mu^+$ on which $\dot{r} \geq 0$; and $\mu^-$ on which $\dot{r} \leq 0$. (This splitting of $\mu$ will not be unique if there exists a non-zero length segment of $\mu$ on which $\dot{r} = 0$. Further, $\mu^-$ will be empty if $\dot{r} > 0$ on $\mu$, while $\mu^+$ will be empty if $\dot{r} < 0$ on $\mu$.) We now argue that the lengths of $\mu^-$ and $\mu^+$ must be less than $3r_M$ thereby establishing theorem 1. It is sufficient to establish this bound on the length of $\mu^-$ as the bound on the length of $\mu^+$ follows by a similar (time-reversed) argument.

Using Eq. (2.5), the non-negativity of $m$, the bound $r \leq r_M$, and the fact that $Q^+Q^- = D^m r D_m r = 1 - 2m/r$, we have

$$\frac{d^2 r}{dt^2} \leq -\frac{1}{2r_M}(1 - Q^+ Q^-). \tag{3.15}$$

It is shown in Appendix B that on $\mu$, and hence on $\mu^-$, the set where $D^a r$ is spacelike must be connected. Therefore, we can and do make that choice of $\epsilon^{ab}$ so that $Q > 0$ and hence $Q^- > 0$ where $D^a r$ is spacelike on $\mu^-$. On the portion of $\mu^-$ where $D^a r$ is future-directed timelike or null or zero, $Q^-$ is necessarily non-negative (independent of the choice of $\epsilon^{ab}$) since $^-k^a$ is past-directed null. Therefore, with this choice,

$$Q^- \geq 0 \text{ on } \mu^-. \tag{3.16}$$

Further, with this choice of $\epsilon^{ab}$, where $D^a r$ is spacelike on $\mu$, $Q^+$ is given by

$$Q^+ = \dot{r} + \sqrt{\dot{r}^2 + 1 - \frac{2m}{r}} \tag{3.17}$$

which is bounded above by 1 on $\mu^-$ since $m \geq 0$ and $\dot{r} \leq 0$. Since $Q^+ \leq 0$ where $D^a r$ is future-directed timelike or null or zero, we have

$$Q^+ \leq 1 \text{ on } \mu^-. \tag{3.18}$$

Using Eqs. (3.15), (3.16), and (3.18) we have on $\mu^-$

$$\frac{d^2 r}{dt^2} \leq -\frac{1}{2r_M}(1 - Q^-). \tag{3.19}$$

Taking the parameterization of $\mu^-$ so that at its past endpoint $t = 0$, integrating Eq. (3.19), and using the fact that $\dot{r}(0) \leq 0$ we have

$$\frac{dr}{dt} \leq -\frac{1}{2r_M}\left(t - \int_0^t Q^-(t')dt'\right). \tag{3.20}$$

Using the bound given by Eq. (3.6c) we then find that

$$\frac{dr}{dt} + \frac{r}{r_M} \leq 1 - \frac{t}{2r_M}. \quad (3.21)$$

Rewriting this inequality in the form

$$\frac{d}{dt}\left(e^{t/r_M} r\right) \leq \left(1 - \frac{t}{2r_M}\right) e^{t/r_M}, \quad (3.22)$$

integrating, and using the fact that $0 \leq r(t) \leq r_M$ (with equality in the lower bound possible only at endpoints of $\mu$) we find that

$$t \leq r_M \left(3 - e^{-t/r_M}\right) < 3r_M, \quad (3.23)$$

showing that the length of $\mu^-$ must be less than $3r_M$. (That timelike curves without endpoints cannot have length $6r_M$ follows from the slightly better bound of $2.95 r_M$ on the lengths of $\mu^-$ and $\mu^+$, which also follows from Eq. (3.23).) This completes the proof of theorem 1.

## IV. DISCUSSION

Can the arguments used in this work be modified so that there is no need for the restrictive condition that $\tau^a{}_a = 0$? After all, in the $k = +1$ Robertson-Walker spacetimes with dust, $D^a D_a r$ is non-positive. For spacetimes such as these, lemma 1 is hopelessly false. Clearly, something other than the quantity $D^a D_a r$ needs to be considered.

As has been emphasized, the bound established for the spacetimes considered here is much simpler than the bounds given in prior works. Is this simplicity peculiar to the spacetimes considered here, or is it an indication of a more general theorem? We conjecture that its beauty is no accident.

**Conjecture:** There exists a number $K$ such that the lifetime of any spherically symmetric spacetime that possesses a compact Cauchy surface $\Sigma$ and that satisfies the dominant-energy and (radial-) non-negative-pressures conditions will be no greater than $K \max_\Sigma (2m)$.

An analysis of the $k = +1$ Robertson-Walker and Kantowski-Sachs spacetimes reveals that this conjecture is satisfied by these spacetimes with $K = \pi$. Furthermore, as the (maximal) dust-filled $k = +1$ Robertson-Walker spacetimes have a lifetime of exactly $\pi \max_\Sigma (2m)$, should such a $K$ exist, it is necessary that $K \geq \pi$.

## APPENDIX A: MAXIMAL RADIAL TIMELIKE CURVES

**Lemma A.** Fix a spherically symmetric globally hyperbolic spacetime $(M, g_{ab})$ and two points $p, q \in M$ with $q \in I^+(p)$. Fix a timelike curve $\mu$ connecting $\mathcal{S}_p$ to $\mathcal{S}_q$ having length $d(\mathcal{S}_p, \mathcal{S}_q)$. Then either: (i) $\mu$ is a segment of $\gamma_n$ or $\gamma_s$; or (ii) none of the points of $\mu$ lying strictly between its endpoints are elements of $\gamma_n$ or $\gamma_s$.

*Proof.* Denote the isometry of $(M, g_{ab})$ corresponding to an element $g \in G$ by $\phi_g$. Suppose that there exists a point $p' \in \mu$ lying strictly between $p$ and $q$ such $r(p') = 0$. If the tangent vector $t^a$ to $\mu$ at $p'$ is invariant under the action of $G$ (i.e., $(\phi_{g*} t)^a = t^a$ for all $g \in G$), then $\mu$ must simply be a portion of $\gamma_n$ (or $\gamma_s$) since $\mu$ is geodetic and its tangent vector is parallel to $\gamma_n$ (or $\gamma_s$). Otherwise, construct the piecewise timelike curve $\mu'$ connecting $\mathcal{S}_p$ to $\mathcal{S}_q$ by taking

$$\mu'(t) = \begin{cases} \mu(t) & \text{for } t_p \leq t \leq t_{p'} \\ \phi_g(\mu(t)) & \text{for } t_{p'} \leq t \leq t_q \end{cases}, \quad (A1)$$

where $g \in G$ is chosen so that $\phi_{g*}$ does not leave $t^a$ fixed. (Note: $\mu(t_p) = p$, $\mu(t_{p'}) = p'$, $\mu(t_q) = q$.) The curve $\mu'$ is continuous, has length equal to the length of $\mu$ being $d(\mathcal{S}_p, \mathcal{S}_q)$, yet its tangent vector is discontinuous at $p'$ (since $t^a \neq (\phi_{g*} t)^a$). It is a standard result that a curve with such a "corner" can always be lengthed by "rounding off the corner" [9]. However, there are no curves from $\mathcal{S}_p$ to $\mathcal{S}_q$ having length greater than $d(\mathcal{S}_p, \mathcal{S}_q)$ showing that in this case such a point $p'$ cannot exist. Therefore, in this case, $r(p') > 0$ for all points $p' \in \mu$ lying strictly between $p$ and $q$. □

## APPENDIX B: CAN'T GO FROM ONE SPACELIKE REGION TO ANOTHER

**Lemma B.** For the spacetimes in theorem 1, if $D^a r$ is future-directed (past-directed) timelike, null, or zero at a point $p$, then $D^a r$ is future-directed (past-directed) timelike on $I^+(p)$ ($I^-(p)$).

*Proof.* Let $p$ be any point at which $D^a r$ is future-directed timelike, null, or zero. (The proof of the past-directed case follows by a similar time-reversed argument.) Fix any point $q \in I^+(p)$ and any radial timelike curve $\gamma$ from $\mathcal{S}_p$ to $q$. Denote the tangent vector to $\gamma$ by $t^a$ and its past endpoint by $p'$. Denote the maximal connected segment of $\gamma$ containing $p'$ on which $m > 0$ by $\gamma'$ and its future endpoint by $q'$. Fixing any radial future-directed timelike or null vector $\xi^a$ at $p'$, extending it to all of $\gamma'$ by parallel transport, then along $\gamma'$ we have

$$\begin{aligned} t^a D_a(\xi^b D_b r) &= t^a \xi^b D_a D_b r \\ &= \frac{m}{r^2}(t^a \xi_a) - \frac{r}{2}(\tau_{ab} t^a \xi^b) \\ &< 0. \end{aligned} \quad (B1)$$

The first equality follows from the fact that $\xi^b$ is parallelly transported along $\gamma'$; the second from Eq. (2.4); and the inequality from the fact that $t^a \xi_a < 0$, the positivity of $m$ on $\gamma'$, and the fact that $\tau_{ab}$ satisfies the dominant-energy condition. Therefore, since $\xi^a D_a r \leq 0$ at $p'$, it must be the case that $\xi^a D_a r < 0$ at $q'$ for all radial future-directed timelike and null vectors $\xi^a$ showing that $D^a r$ must be future-directed timelike at $q'$. From this we conclude that $q'$ must in fact be $q$. Therefore, $D^a r$ is future-directed timelike at $q$ as was to be shown. □

(Lemma B is false for more general spherically symmetric spacetimes. A simple counterexample is provided by a $k = +1$ Robertson-Walker spacetime with dust as a source.)

It follows from lemma B that the portion of any timelike curve $\gamma$ where $D^a r$ is spacelike must be connected since this segment is simply $\gamma$ less that portion of $\gamma$ on which $D^a r$ is future-directed timelike, null, or zero (which, by lemma B, is is either all of $\gamma$ or the portion of $\gamma$ lying to the future of some point of $\gamma$) and that portion of $\gamma$ on which $D^a r$ is past-directed timelike, null, or zero (which again, by lemma B, is either all of $\gamma$ or the portion of $\gamma$ lying to the past of some point of $\gamma$).

A simple consequence of this result is that it is impossible for an observer to travel from $\gamma_n$ to $\gamma_s$ (or vice-versa).